\documentclass[fleqn,usenatbib,letter]{aa}

\usepackage{graphicx}
\usepackage{txfonts}
\newcommand{\cms}{\mbox{cm\,s$^{-1}$}}
\newcommand{\muhz}{\mbox{$\mu$Hz}}

\begin{document}

\title{Oscillations in the Sun with SONG: Setting the scale for asteroseismic investigations\thanks{Based on observations made at the Hertzsprung SONG telescope operated at the Spanish Observatorio del Teide on the island of Tenerife by the Aarhus and Copenhagen Universities and by the Instituto de Astrofísica de Canarias.}}

   \author{M.~Fredslund~Andersen\inst{1}
          \and
          P.~Pall\'e\inst{2,3}
          \and
          J.~Jessen-Hansen\inst{1}
		  \and
		  K.~Wang\inst{2,4}
		  \and
		  F.~Grundahl\inst{1}
		  \and
		  T.~R.~Bedding\inst{5,1}
		  \and
		  T.~Roca~Cortes\inst{2,3}
		  \and
		  J.~Yu\inst{5,1}
		  \and
		  S.~Mathur\inst{2,3}
		  \and
		  R.~A.~Gacia\inst{6,7}
		  \and
		  T.~Arentoft\inst{1}
		  \and
		  C.~Régulo\inst{2,3}
		  \and
		  R.~Tronsgaard\inst{8}
  		  \and  
          H.~Kjeldsen\inst{1}
          \and
          J.~Christensen-Dalsgaard\inst{1}
          }

\institute{Stellar Astrophysics Centre, Aarhus University, DK-8000 Aarhus C, Denmark
\and
Instituto de Astrofísica de Canarias. E-38205 La Laguna, Tenerife, Spain 
\and
Universidad de La Laguna (ULL), Departamento de Astrofísica, E-38206 La Laguna, Tenerife, Spain
\and
Department of Astronomy, China West Normal University, No 1 Shi Da Road, Nanchong City, Sichuan Province, China
\and
Sydney Institute for Astronomy, School of Physics, University of Sydney, Camperdown New South Wales 2006, Australia
\and
Université Paris Diderot, AIM, Sorbonne Paris Cité, CEA, CNRS, F-91191 Gif-sur-Yvette, France
\and
IRFU, CEA, Université Paris-Saclay, F-91191 Gif-sur-Yvette, France
\and
DTU Space, National Space Institute, Technical University of Denmark, Elektrovej 328, DK-2800 Kgs. Lyngby, Denmark
}

   \date{Received January 31, 2019; accepted February 25, 2019}

 
  \abstract
   {We present the first high-cadence multi-wavelength radial-velocity observations of the Sun-as-a-star, carried out during 57 consecutive days using the stellar échelle spectrograph at the Hertzsprung SONG Telescope operating at the Teide Observatory.}
   {The aim was to produce a high-quality data set and reference values for the global helioseismic parameters $\nu_{\rm max,\odot}$ and $\Delta \nu_{\odot}$ of the solar p-modes using the SONG instrument. The obtained data set or the inferred values should then be used when the scaling relations are applied to other stars showing solar-like oscillations which are observed with SONG or similar instruments.}
   {We used different approaches to analyse the power spectrum of the time series to determine $\nu_{\rm max,\odot}$; simple Gaussian fitting and heavy smoothing of the power spectrum. $\Delta \nu_{\odot}$ was determined using the method of autocorrelation of the power spectrum. The amplitude per radial mode was determined using the method described in \citet{kjeldsen-2008}.}
   {We found the following values for the solar oscillations using the SONG spectrograph: $\nu_{\rm max,\odot}=3141\pm12\,\mu$Hz, $\Delta \nu_{\odot}=134.98 \pm 0.04\,\mu$Hz and an average amplitude of the strongest radial modes of $16.6 \pm 0.4 \,\cms$.  These values are consistent with previous measurements with other techniques. }
   {}

   \keywords{The Sun --
                Solar Oscillations --
                Asteroseismology}

   \maketitle
%

\section{Introduction}
The Stellar Observations Network Group, SONG, aims at building a network of 1-m telescopes well spread in longitude, in both the northern and the southern hemispheres. Each node in the network is designed to be fully automatic, with no human interactions needed on site \citep{mfa2014,mfa2019}. One of the primary goals of SONG is to target individual stars intensively for asteroseismic studies using high-precision radial-velocity measurements \citep[e.g.:][]{fgj2006,fgj2014,fgj2017}. Furthermore, with the complementary satellite missions such as \textit{TESS} \citep{tess}, simultaneous observations combining space based photometry and radial velocities from SONG provide new and valuable insight in our understanding of stars.

Global seismic parameters obtained from observations yield fundamental global properties of stars, like mass and radius, using well-established scaling relations \citep{brown1991,kjeldsen-bedding1995,stello2009,kallinger2010}. These are based on scaling two key properties of pressure mode (p-mode) oscillations from the Sun to other stars, namely the large frequency separation ($\Delta \nu$) and the frequency of maximum power ($\nu_{\rm max}$).
While $\Delta \nu$ should be largely independent of the instrument used, the value of $\nu_{\rm max}$ is sensitive to the depth of the sounded layers, which depends on the observational technique. A number of radial-velocity observations have been carried out for the Sun using several ground-based networks (BiSON, GONG, and now SONG) and space missions (GOLF/SoHO and HMI/SDO), which are summarized in \citet{2018Pere}. Each instrument is looking at a specific part of the solar spectrum to determine the radial velocities and so it is sensitive to different depths in the solar atmosphere; the profile of the p-mode power envelope depends on the response function of every spectral line as a function of the height in the atmosphere. Hence different profiles of the envelope are observed when using different monochromatic instruments where single spectral lines are used (i.e. K-7699$\AA$ for BiSON, Ni-6768$\AA$ for GONG etc.). This will translate into differences in the measured properties of the oscillations. Using the SONG spectrograph and the iodine technique where many spectral lines are used to extract the velocities will result in an ``average'' measurement insensitive to effects originating at specific heights in the solar atmosphere. Therefore the observations of the Sun, presented here, are very important for applying the scaling relations when oscillations are observed in other stars using the same (or a similar) instrument, in order to minimize systematic effects.

As shown in \citet{palle2013}, simultaneous observations of the Sun using different instruments (including the SONG spectrograph) result in different profiles of the oscillation envelope and hence in different values of $\nu_{\rm max}$. In addition, the stellar background from granulation is less dominant in velocity than in intensity \citep[e.g.,][]{palle1999,kjeldsen-bedding2011}. In this letter, we emphasize the importance of comparing stellar values determined from SONG data to our observations of the Sun using SONG. The obtained time series and power spectrum are available from the SONG Data Archive\footnote{http://soda.phys.au.dk} for future use.

\section{Observations and data reduction}
The Hertzsprung SONG telescope at the Teide Observatory on the island of Tenerife has been operating in scientific mode since March 2014 \citep{mfa2016}. It has observed many asteroseismic targets using the high-resolution \'echelle spectrograph for radial-velocity measurements \citep[e.g.:][]{fgj2017,stello2017,frandsen2018,arentoft2019}. 

\begin{figure}
\includegraphics[width=\columnwidth]{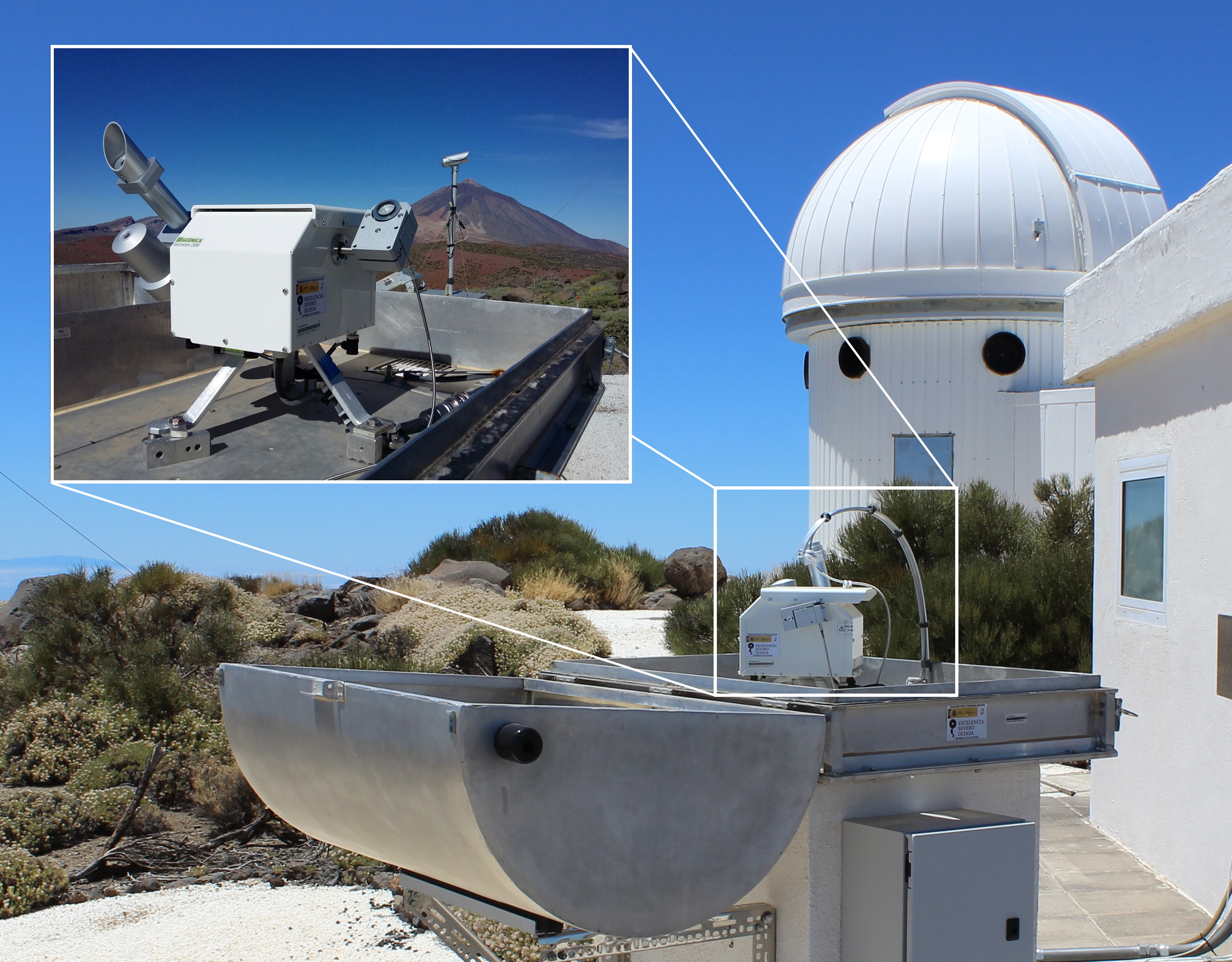}
\caption{The solar tracker with the attached optical fibre assembly used to feed sunlight to the SONG spectrograph. The SONG dome and container are seen in the background.}
\label{fig:solar-tracker}
\end{figure}

In 2017, the so-called ``Solar-SONG'' complementary initiative was funded and a solar tracker was installed next to the SONG telescope. This allows light from the Sun to be fed directly into the SONG spectrograph using an optical fibre assembly with complementary optics to scramble the sunlight \citep{fibre}.
The optical fibre is mounted on a dedicated Alt/Az Solar tracker and pointed directly towards the Sun (see Fig.~\ref{fig:solar-tracker}). The tracker has the fibre mounted on one side, and a pyrheliometer and an active guide unit on the other side. With this instrument we could simultaneously collect the total solar irradiance (TSI) and the solar spectra. The TSI was used to clean the data for bad points, primarily caused by clouds. 

A diffuser was placed at the fibre entrance which, together with ball lenses at the interface to the octagonal and circular fibres, was intended to ensure that the solar disk was not resolved, even if the tracker does not point accurately to the same point on the Sun at all times. The active guide unit was not functioning for the observations presented in this paper and, as discussed below, some effects of the solar disk being resolved were seen in the data. 

The observations were carried out from May 27 to July 22 2018 corresponding to an extremely deep minimum of the solar activity cycle. Activity will affect the solar oscillations by lowering the p-mode amplitudes and therefore observing during an activity minimum is highly favorable. Each day more than 10 hours of data were collected where each spectrum had an exposure time of 0.5 seconds and a readout time of 3.5 seconds, resulting in approximately 12,000 spectra (about 100\,GB) per day and a total of more than 500,000 measurements of radial velocities after removing bad points. The extraction of the 2-D spectrum into the flux-wavelength spectrum was set to work in real time. The {\tt iSONG} pipeline was used to produce the radial-velocity values \citep{corsaro2012,antoci2013} and was set up on a computer cluster at the Instituto de Astrofísica de Canarias on Tenerife where the reduction tasks were handled by a HTCondor distributor.

\begin{figure}
\includegraphics[width=\columnwidth]{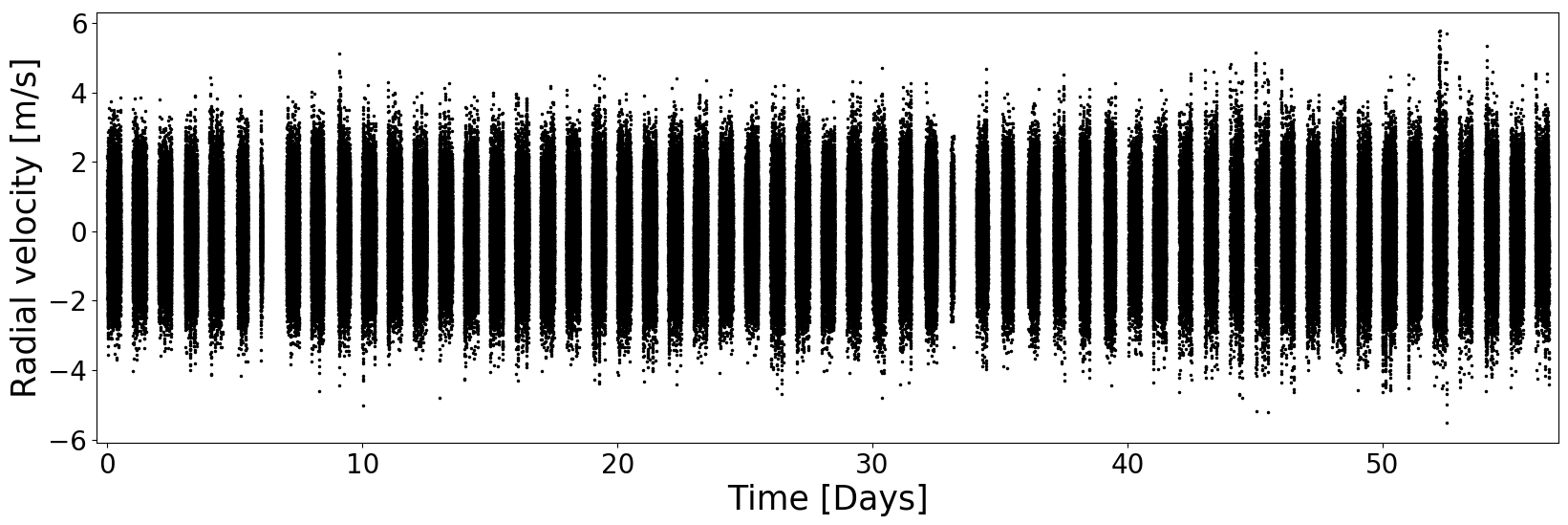}
\includegraphics[width=\columnwidth]{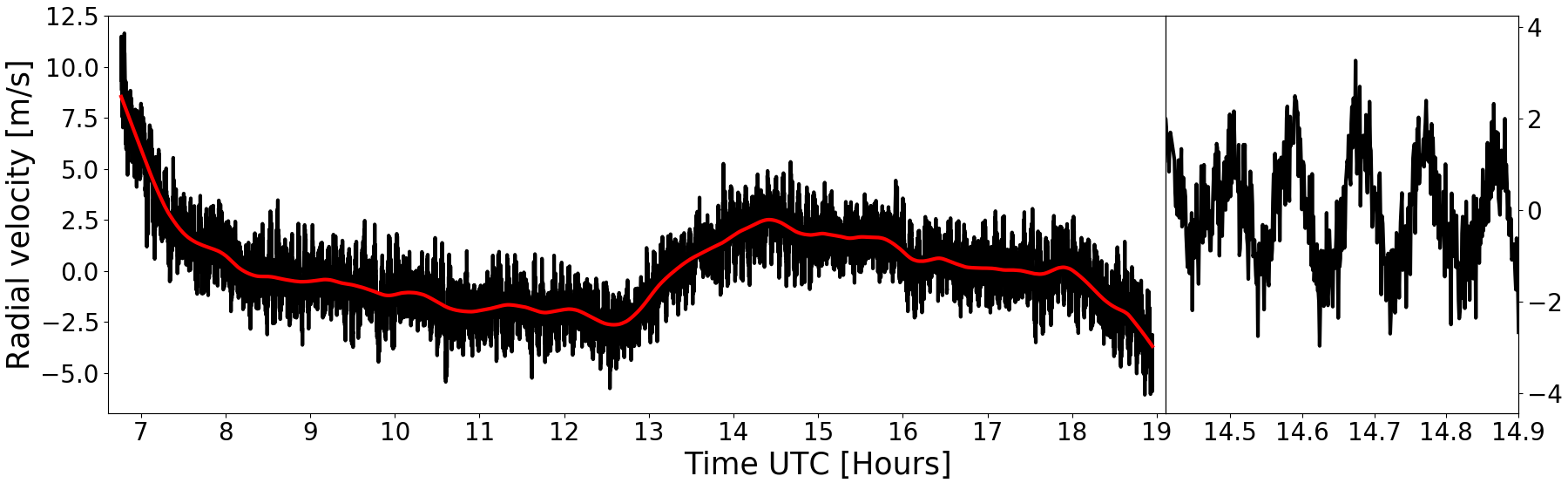}

\caption{Upper: The full 57 days solar time series after corrections and filtering have been applied.
\label{fig:time_series_full}
Lower left: The barycentric velocity corrected time series from one day of solar observations. Some trends in the residual RV curve are clearly visible. The red line is the used local regression smoothing filter. Lower right: A zoom on the filtered and corrected time series, in which the 5-minute oscillations are clearly seen.
\label{fig:time_series_raw}}
\label{fig:time_series_cor}
\end{figure}

Fig.~\ref{fig:time_series_full} (upper panel) shows the full radial-velocity time series after corrections and filtering have been applied. Only two of the 57 days in this campaign had significant interruptions resulting in a duty cycle of $\sim$40\%. On June~2 the tracker failed after two hours of observations due to a software hang-up, and on June~29 clouds were covering the observatory. The raw radial-velocity measurements for each day were first corrected for the barycentric motion of the Earth around the Sun (\citet{barycor} code, exported to Python). The residual time series of one full day is shown in Fig.~\ref{fig:time_series_raw} (lower left panel), where some instrumental effects are still present. The trend near noon originates from the tracker not pointing perfectly to the same point on the Sun at all times. This is also where the largest effects are expected when using an Alt/Az mount. The observations are consistent with the fact that the solar disk is partially resolved, so that the setup does not completely observe the Sun as a star. The downward slope in the residuals at the beginning and end of the series is a well known effect \citep{Belmonte1988} and is caused by differential extinction in the Earths atmosphere. The low-frequency trends were removed using local weighted linear regression smoothing (LOWESS) \citep{lowess}, which is seen as the red curve on Fig.~\ref{fig:time_series_raw} (lower left panel). 
Each data point was smoothed using a weighted linear regression on a subset of the radial-velocity measurements. The subset used around the individual points was specified as a fraction of the total number of data points and in our case we chose a fraction value of 0.05.
With roughly 12.000 data points per day this corresponds to a high-pass filter with cutoff frequency close to 450 $\mu$Hz. A number of different filters were tested in order to choose one that had no effect on our main results on the p-mode oscillations.
Each filter was checked, after being applied to the barycentric velocity corrected time series, by determining $\nu_{max}$ of the p-mode power excess and the differences between the filters were within a few $\mu$Hz. Finally, the filter that minimized the morning, noon and evening trends was chosen.
Fig.~\ref{fig:time_series_cor} (lower right panel) shows a zoom of the corrected and filtered time series for one day where the solar oscillations are clearly visible.

\section{Analysis of oscillations}
\begin{figure}
\includegraphics[width=\columnwidth]{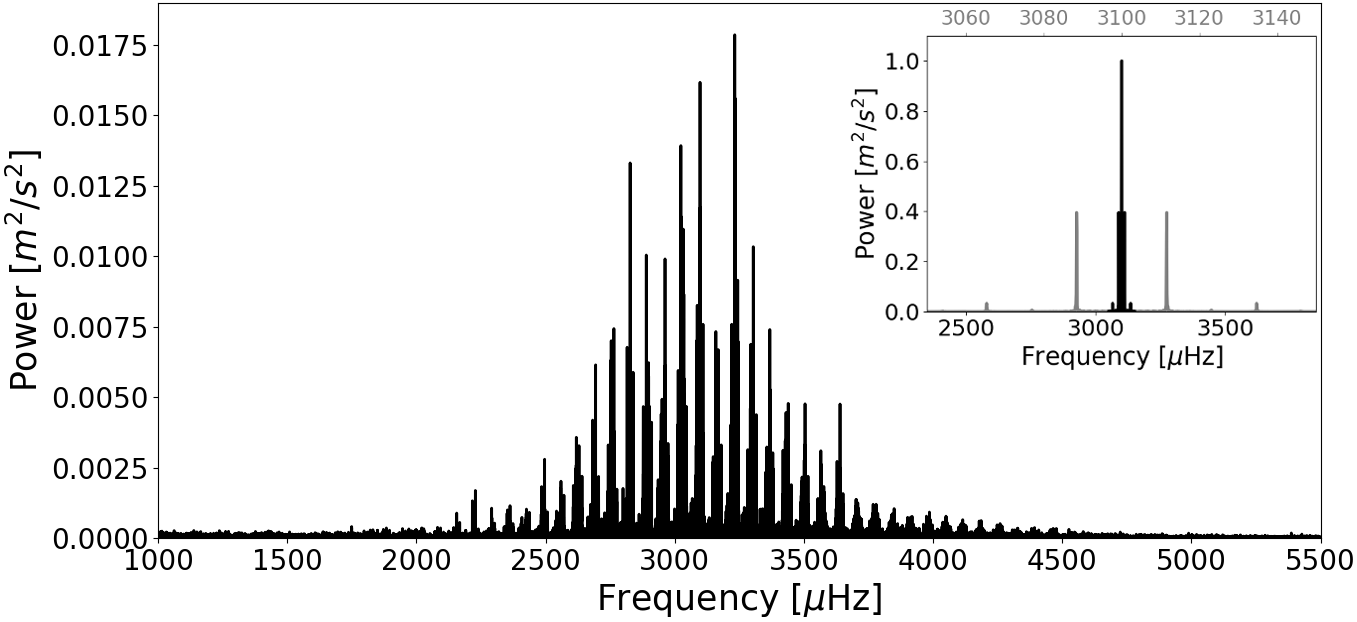}
\includegraphics[width=\columnwidth]{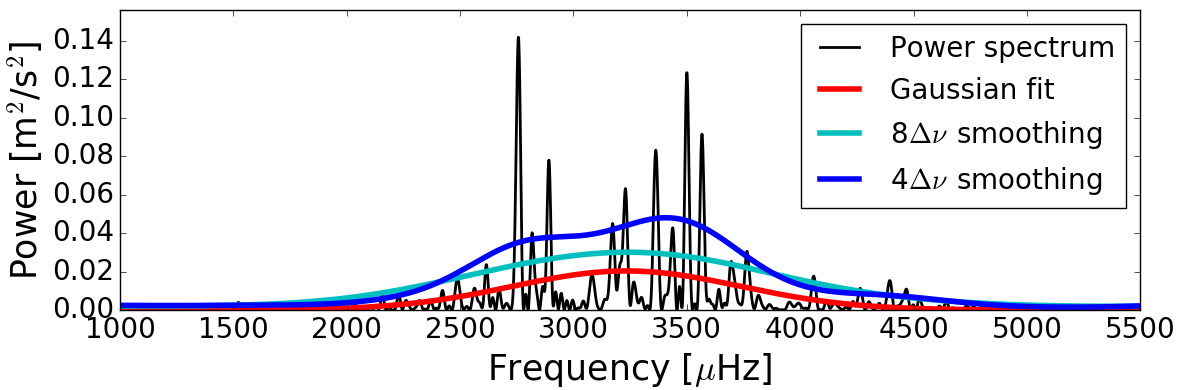}
\caption{Upper: The power spectrum from the full 57 days filtered and corrected time series. The inset (black) shows the spectral window ($\sim$40\% duty cycle) and (gray) a zoom to show the individual aliases.
\label{fig:full_ps}
Lower: The power spectrum of one single day (black). The red curve is a Gaussian fit to the power excess, the cyan curve is a $8\Delta\nu$ smoothed version of the power spectrum and the blue is the $4\Delta\nu$ smoothed curve. For this day, the $4\Delta \nu$ smoothing results in a double hump, giving a poorly determined $\nu_{\rm max}$. To better distinguish the three colored curves in the lower panel the $8\Delta\nu$ and the $4\Delta\nu$ smoothed versions were multiplied by a factor of 2 and 2.5 respectively.}
\label{fig:1d-nu_max}
\end{figure}

To determine the global helioseismic parameters, power spectra of the corrected and filtered time series were calculated using unweighted iterative least-squares sine-wave fitting. The power spectrum from the full 57 days is shown in Fig.~\ref{fig:full_ps} (upper panel). 

\subsection{Determination of $\nu_{\rm max}$}
\begin{figure}
\includegraphics[width=\columnwidth]{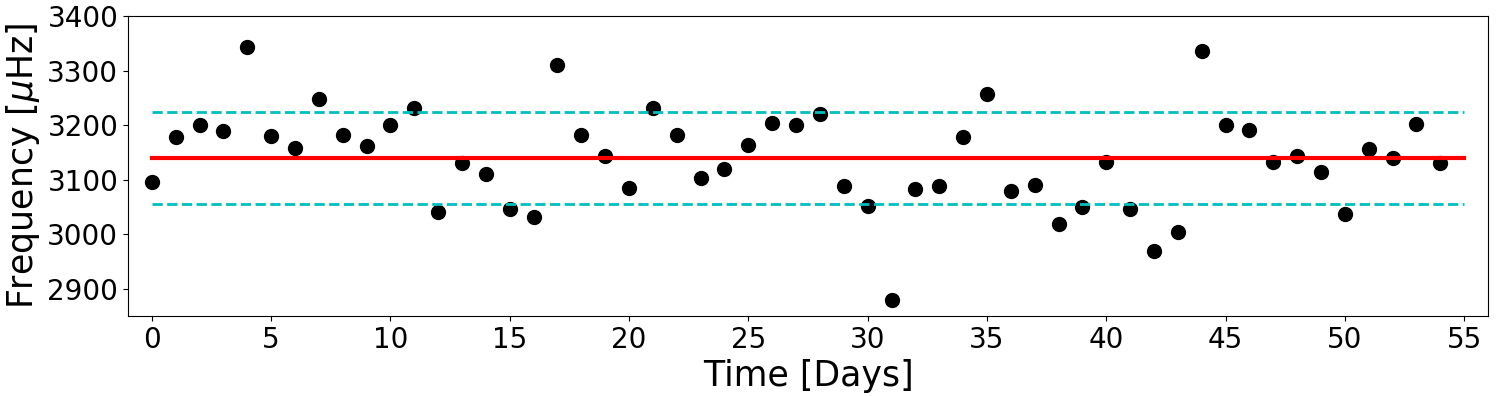}
\caption{The $\nu_{\rm max}$ time series showing the values determined on the single-day power spectra using a Gaussian fit. 
The large variations are due to the stochastic nature of the modes where excitation, damping and interactions causes the heights of individual modes to vary with time. The red line is the mean value and the dashed cyan lines show $\pm 1\sigma$. 
}
\label{fig:nu_max_ts}
\end{figure}

To estimate uncertainties on the helioseismic parameters, the time series were split into chunks of one day each. In one day, the 5-minute oscillations still produces a significant signal in the power spectrum ($S/N \sim 40$), which can be seen in Fig.~\ref{fig:1d-nu_max} (lower panel). 
Different methods were applied to determine the frequency of maximum power ($\nu_{\rm max}$) from the individual power spectra.
One was to fit a simple Gaussian to the power excess. The other method was to apply the procedure of \citet{kjeldsen-2008}, where the power spectrum is smoothed by a Gaussian with a FWHM of $4\Delta \nu$ and $\nu_{\rm max}$ determined as the maximum of the smoothed curve. The smoothing factor when using the second method proved to be too low and the value was increased to a FWHM of $8\Delta \nu$. The lower value would in some cases result in a multi hump which would lead to a badly determined $\nu_{\rm max}$ (see lower panel of Fig.~\ref{fig:1d-nu_max}).\\ 
Some tests using different background models, when applying the Gaussian fit, were tried with effects well below the uncertainty and we therefore decided to omit a background term.   
With these two methods we got a series of independent measurements of $\nu_{\rm max}$ each day for the Sun (see Fig.~\ref{fig:nu_max_ts}). Calculating the standard deviation ($\sigma$) of these values gives a standard mean error of $\sigma/\sqrt{n}$, where $n=55$ is the number of independent determinations (days used). The two days with only a few data points were omitted. The determined uncertainties were: $\pm 12 \,\muhz$ for the Gaussian method and $\pm 11 \,\muhz$ using the $8\Delta \nu$ smoothing method. We adopt the value from the Gaussian method from now on.

\noindent We then measured $\nu_{\rm max,\odot}$ by fitting a Gaussian to the power excess of the full power spectrum and the value was: 
\begin{eqnarray}
\nu_{\rm max,\odot} = 3141 \pm 12 \,\muhz\quad 
\label{nu_max}
\end{eqnarray}
The mean values from the 55 independent measurements of $\nu_{\rm max}$ results in similar values for the two methods: $\nu_{\rm max,\odot} = 3139\pm 12 \,\muhz$ (Gaussian - shown in Fig.~\ref{fig:nu_max_ts}) and $\nu_{\rm max,\odot} = 3140\pm 11 \,\muhz$ ($8\Delta \nu$ smoothing). We also used two widely used pipelines to extract both $\nu_{\rm max}$ and $\Delta\nu$: SYD Pipeline \citep{huber2009}: $\nu_{\rm max,\odot} = 3141 \pm 18 \,\muhz$ and the \textit{A2Z} Pipeline \citep{A2Z}: $\nu_{\rm max,\odot} = 3151 \pm 147 \,\muhz$, which all agrees within the uncertainties.

\subsection{Determination of $\Delta\nu$}
To determine $\Delta\nu$, we used the method of calculating the autocorrelation of the full power spectrum.  We smoothed the power spectrum slightly before calculating the autocorrelation and determined the value by fitting a Gaussian to the peak identified as originating from $\Delta\nu$. The value with the standard error of the Gaussian fit was
\begin{equation}
 \Delta\nu_{\odot} = 134.98 \pm 0.04\,\muhz,
\end{equation}
which is in agreement with the literature \citep[e.g.:][]{golf_delta, kjeldsen-2008}. The results from the pipelines were: SYD Pipeline: $\Delta\nu_{\odot} = 135.02 \pm 0.09\,\muhz$ and the \textit{A2Z} Pipeline: $\Delta\nu_{\odot} = 134.81 \pm 3.11 \,\muhz$, which all agrees well.

\begin{figure}
\includegraphics[width=\columnwidth]{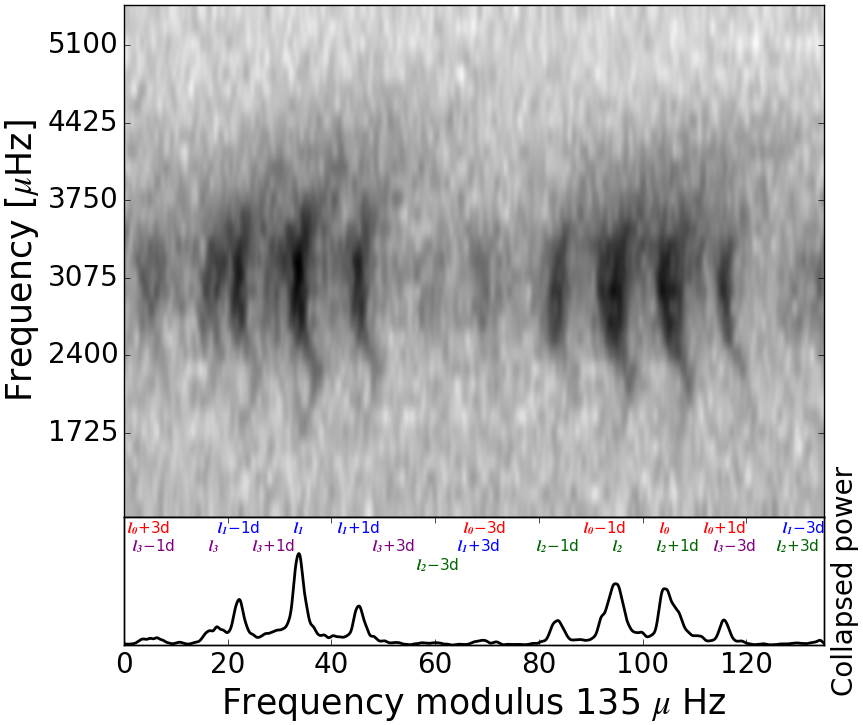}
\caption{(Top panel) The échelle diagram of a slightly smoothed version of the power spectrum. The ridges from the p-modes with different $l$-value are clearly visible as well as the alias peaks from the window function. (Bottom panel) The collapsed échelle diagram with labels centered above the peaks marking the $l_{\{0,1,2,3\}}$ ridges and the $\pm$1 and $\pm$3 daily aliases (no 2 day alias present).}
\label{fig:echelle}
\end{figure}

The determined value of $\Delta\nu$ was evaluated further by creating an échelle diagram where the slightly smoothed power spectrum was cut into chunks with a length of $\Delta\nu$ and placed on top of each other. With the correct value of $\Delta\nu$ vertical ridges originating from modes with different $l$-values and the corresponding daily aliases will appear. The échelle diagram of the power spectrum is seen in Fig.~\ref{fig:echelle}.

\subsection{Oscillation amplitude}
The amplitude of the solar oscillations was determined with the SYD pipeline, which uses the procedure described by \citet{kjeldsen-2008}. This involves first converting the power spectrum of the full 57 days into power spectral density (PSD) by multiplying by the effective observing time (22.6\,d), which we calculated as the reciprocal of the area under the spectral window. The PSD was then heavily smoothed by a Gaussian with a FWHM of $4\Delta\nu_{\odot}$ and multiplied by $\Delta\nu_{\odot}/c$, where we took $c=4.09$, which represents the effective number of modes in each order \citep{kjeldsen-2008}. We then fitted and subtracted the background using a two-component Harvey model \citep{harvey1985}. The combination of the components in the model were equivalent to a linear fit. Finally, we converted to amplitude by taking the square root and found the amplitude measured in this way, which is the amplitude per radial mode, to be
\begin{equation}
 A_\odot = 16.6 \pm 0.4 \,\cms.  
 \label{amp}
\end{equation}
Using the same method but a different instrument (BiSON), \citet{kjeldsen-2008} found the long-term (11 years) average of the solar amplitude to be $18.7 \pm 0.7$\cms, with significant scatter over time due to the stochastic nature of the modes and solar activity. With this in mind, our measurement appears to be consistent with the previous result. The oscillation amplitudes will generally be affected by the window function which will lower the amplitudes \citep{arentoft2019}. Simulations were performed to check the effect on our data set and the effect was below the level of the stated uncertainty in (\ref{amp}).

\section{Conclusion}
We have presented the first multi-wavelength high-cadence radial-velocity observations of the Sun-as-a-star to date, using the SONG spectrograph on Tenerife. We applied standard methods to determine the global helioseismic values $\nu_{\rm max,\odot}$ and $\Delta \nu_{\odot}$. 
The value of $\nu_{\rm max,\odot} = 3141 \pm 12\,\muhz$ determined here shows one way of determining $\nu_{\rm max}$ when analysing SONG data. 
Our value is an average over different depths in the solar atmosphere which makes it (data and method) a good reference for future use in the stellar scaling relations. The method of fitting a Gaussian to the p-mode envelope of the daily power spectra and to the full power spectrum to determine $\nu_{\rm max}$ and its associated error can directly be applied to all other asteroseismic targets observed using SONG and will lead to a homogeneous and robust way of determining $\nu_{\rm max}$ with a realistic uncertainty. 
We determined $\Delta \nu_{\odot}$ of the Sun using autocorrelation to be $134.98 \pm 0.04\,\muhz$.
The value of $\nu_{\rm max,\odot}$ and $\Delta \nu_{\odot}$ determined here also confirms the instrument performance and pipeline for the radial-velocity measurements of SONG. Finally, we found the amplitude of the strongest radial modes to be $16.6 \pm 0.4\,\cms$, which is consistent with previous measurements.  
These values, especially $\nu_{\rm max,\odot}$, will be very important when the scaling relations are applied to other stars showing solar-like oscillations observed with SONG or similar instruments.
In case other methods will be applied to extract the global asteroseismic values of SONG targets we have made the filtered and corrected time series and corresponding power spectrum available on the SONG Data Archive (SODA).

\section*{Acknowledgements}
We would like to acknowledge the contribution of the engineers at the IAC Felix Gracia (optical), Ezequiel Ballesteros (electronic) and Antonio Dorta (HTCondor) and the support from the day operator at the SolarLab, Teide Observatory, on this Solar-SONG initiative and the campaign. We would also like to acknowledge the Villum Foundation, the Independent Research Fund Denmark and the Carlsberg Foundation for the support on building the SONG prototype on Tenerife. The Stellar Astrophysics Centre is funded by The Danish National Research Foundation (Grant DNRF106). Funding for the “Solar-SONG” provided by the Excellence “Severo Ochoa” program at the IAC and the Ministry MINECO under the program AYA-2016-76378-P. R.A.G. is supported by the GOLF/SOHO grant by the CNES. S.M. acknowledges support from the Ramon y Cajal fellowship number RYC-2015-17697.

\bibliographystyle{aa} 
\bibliography{sun}

\twocolumn

\end{document}